# The orbital period of the eclipsing dwarf nova CG Draconis

Jeremy Shears, David Boyd, Steve Brady & Roger Pickard

## Abstract

We have performed time resolved photometry on the dwarf nova CG Dra and have established for the first time that it is an eclipsing system. By measuring the times of the eclipses, we determined the orbital period as 0.18864(4) d, or 4h 31m 38 +/- 3s. This value is consistent with the shorter of two periods proposed from earlier spectroscopic studies. The orbital period places CG Dra above the period gap. The symmetrical eclipses are of short duration (FWHM 18+/-2 min, or 0.066(7) of the orbital period) and shallow (average 0.16+/-0.02 mag), suggesting a grazing eclipse which is consistent with an orbital inclination just above the critical value. Flickering persists through the eclipse which means that the flickering source is not occulted by the secondary star.

## Background

CG Draconis was discovered by Hoffmeister in 1965 [1] and he later classified it as a dwarf nova [2]. Rather little was known about its outburst frequency until an intensive monitoring campaign was conducted by two of the authors (JS and RP) and Gary Poyner during 2005, the results of which were published in an earlier edition of the Journal [3]. Their observations, combined with others from the BAA and AAVSO databases showed that the star in fact shows frequent outbursts with a period of around 11 days and the star spends very little time at quiescence. Two types of outburst were detected: short outbursts lasting about 4 days and long outbursts lasting about 8 days.

Dwarf novae are binary stars comprising a cool main sequence star (the secondary star) and a white dwarf primary. Matter flows from the secondary towards the white dwarf and forms an accretion disc around the white dwarf. As material builds up in the disc, a thermal instability is triggered that drives the disc into a hotter, brighter state causing an outburst in which the star brightens, sometimes by several magnitudes [4]. A spectroscopic study of CG Dra by Bruch *et al.* revealed two spectral components, which are attributed to the accretion disc and the secondary, and whose characteristics are consistent with it being a dwarf nova [5]. However, there are a number of peculiarities in the spectrum which are difficult to explain by the standard models for dwarf novae and Bruch *et al.* could not provide a self-consistent description of the system. Their radial velocity measurements indicated an orbital period of either 0.1893(6) d (4h 32min +/- 1 min) or 0.2343(21) d (5h 37min +/-3min).

JS and RP carried out several short time series photometry runs on CG Dra in 2005, during which flickering was obvious. Small non-periodic modulations of up to 0.14 mag were also detected by Tonny Vanmunster during photometry conducted in 2005 May [6]. There was also the suggestion of some intriguing dips in the light curves produced by these observers. However, the results were inconclusive and it was not known whether the dips were random or regular features, hence a campaign was organised by the authors in 2007 to carry out more extensive photometry, the results of which are discussed in this paper.

**Time resolved photometry**

Unfiltered time resolved photometry was conducted using the instrumentation described in Table 1. Integration times were 60 s for JS, RP and DB and 30 s for SB. A total of 15 separate photometry runs were conducted, according to the observing log in Table 2, yielding more than 54 h of data. In all cases raw images were flat-fielded and dark-subtracted, before being analysed using commercially available differential aperture photometry software against the comparison star sequence given in the AAVSO f-chart for CG Dra dated 020301 [7]. An image of the star in outburst is shown in Figure 1.

**Detection of eclipses and measurement of the orbital period**

Examples of four of the photometry runs are shown in Figures 2 to 5, where dips are clearly present in the light curve which, because of their periodicity, we interpret as eclipses. During the campaign a total of 13 such eclipses were observed. The times of minimum of each eclipse were determined according to the Kwee & van Woerden method [8] using the Peranso software [9]. These are listed in Table 3 where the eclipses are labeled with the corresponding orbit number starting from 0. The orbital period was then calculated by a linear fit to these times of minima as $P_{orb}$ = 0.18864(4) or 4h 31m 38 +/- 3s. The eclipse time of minimum ephemeris is:

$$JD_{min} = 2454360.41287(3) + 0.18864(4)*E$$

The O-C (Observed – Calculated) residuals of the eclipse minima relative to this ephemeris, given as a fraction of the orbital period, are shown in Figure 6. This confirms that the period remained constant during the 26 days (116 orbits) spanned by the observations. However, we note that there is some scatter in O-C values which is probably due to the difficulty in isolating eclipse minima relative to flickering in the light curve.

We should also point out that 5 of the time series photometry runs encompassed more than a complete orbital period, during which we saw no additional eclipses, other than those expected from the above ephemeris. Furthermore, we found no cases where an eclipse which was predicted by our ephemeris was not in fact detected.

We note that our value of $P_{orb}$ is consistent with the shorter of the two periods proposed by Bruch *et al.* from radial velocity measurements [5] and certainly rules out their second, longer period. It places CG Dra well above the "period gap" in the orbital period distribution of dwarf novae [4]. It is unlikely that CG Dra is a member of the UGZ family of dwarf novae, although this has never been suggested, on the basis that no standstills have been detected [3] in which the star remains between outburst and minimum for long periods and which are characteristic of this family. However, as the star has only been monitored for a relatively short time, it is possible that standstills have yet to be detected and we encourage further monitoring. This, combined with its very frequent outbursts [3], suggests that CG Dra is likely a member of the UGSS family.

## The nature of the eclipses

Eclipses occur in binary systems, whenever the observer's line of sight lies close to the orbital plane (i.e. the orbital inclination is sufficiently high) such that the stars periodically block out light from each other. In the case of dwarf novae, the accretion disc, which varies in size according to whether the system is in outburst, presents an extended source of light to be eclipsed by the secondary. Hence timing of eclipses can provide useful information about the state of the accretion disc. Careful studies of the eclipse profile can yield information about the structure of the system, including the shape and size of the accretion disc, the inclination of the system, *i*, and the location of the white dwarf and the bright spot, where the material flowing from the secondary impacts the accretion disc. A review of eclipsing dwarf novae, and how eclipses can be used to probe their structure, was recently presented in this Journal by Worraker and James [10].

Inspection of the eclipses of CG Dra in Figures 2 to 5 shows that the eclipse profiles are almost symmetrical, indicating an axisymmetric brightness distribution in the accretion disc. Analysis of the eclipse times was complicated by the large amount of flickering shown by the star of up to 0.1 mag, which made it difficult to determine exactly when the eclipse started and finished. We determined the width of each eclipse by drawing a baseline corresponding to the average lightcurve before and after the eclipse and then measured the full width at half minimum (FWHM) of each eclipse; given the scatter due to flickering there are errors of +/-2 min on this value. Table 3 shows that the eclipse FWHM ranged between 15+/-2 min and 20+/-2 min with an average of 18+/-2 min. The short duration of the eclipse, which amounts to only 0.066(7) of the orbital period, i.e. $\Delta\Phi_{1/2} = 0.066(7)$, suggests that the inclination of the system is close to the critical value, below which eclipses do not occur, resulting in a grazing eclipse. The average critical value for dwarf novae is about 71$^{\circ}$, but the limiting value for an individual system depends on the relative size of the secondary star [10]. Our value of $\Delta\Phi_{1/2}$ for CG Dra is lower than that for the eclipsing dwarf nova IP Peg ($\Delta\Phi_{1/2} \approx 0.09$ in outburst), which has $i \approx 81^{\circ}$ [10] and similar to that of HS 0907+1907 where $\Delta\Phi_{1/2} \approx 0.06$ and $i \approx 73\text{-}79^{\circ}$ [11].

The duration of the eclipse is apparently independent of the outburst status of the star since our study includes observations both in outburst (when the star is ~ mag 16, e.g. Figures 2 and 5) and quiescence (e.g. ~ mag 16.7, e.g. Figures 3 and 4). In dwarf novae where there are deep eclipses, it is commonly found that the eclipse duration reduces as the star declines from outburst, because the accretion disc is largest at the peak of the outburst and thereafter shrinks [12, 13]. The fact that we did not see this in the case of CG Dra could mean that the width of the accretion disc varied little between outburst and quiescence, which is possible since the outburst amplitude of CG Dra is rather modest compared with many dwarf novae where magnitude ranges of 2 or more are found. CG Dra is similar in this respect to the low amplitude dwarf nova V729 Sgr whose outburst range is typically ~ 1 mag, but occasionally ~1.5 mag and where $\Delta\Phi_{1/2}$ is largely unchanged between high and low states at $\Delta\Phi_{1/2} = 0.15$ [14]. Whilst the actual value of *i* has yet to be determined for V729 Sgr, it appears to be just above the critical value [14].

We also measured the depth of the eclipses, Δm, but again this was complicated by the flickering, which results in an error of +/-0.02 mag on each measurement. The data in Table 3 show that Δm was found to be between 0.14 +/- 0.02 mag and 0.21+/-0.02 mag, with the average Δm = 0.16+/- 0.02 mag. Such value of Δm is rather modest compared to some eclipsing dwarf novae, but would be expected from a system whose inclination is just above the critical value, resulting in only a partial eclipse of the accretion disc and not the white dwarf. Whilst the value of Δm appears to be independent of whether CG Dra was in outburst or not, we cannot be definitive about this given the inherent errors in each measurement. High inclination, deeply eclipsing, dwarf novae tend to show much deeper eclipses in quiescence as the principle light sources, the white dwarf and the bright spot, are occulted by the secondary. By contrast, low inclination eclipsing systems show much less variation between outbursts and quiescence; for example V729 Sgr, which was mentioned above, has Δm ≈ 0.2 mag in outburst and 0.3-0.5 mag in a low state [14]. Hence our observed values of Δm are again consistent with CG Dra having a relatively low inclination.

Inspection of the lightcurves shows that flickering continues throughout the eclipse, i.e. the source of flickering is not fully occulted. Flickering in dwarf novae is thought to arise from two sources: one is the inner disc, close to the white dwarf, and the other is the bright spot [15]. If CG Dra is indeed a low inclination system, it is entirely possible that the inner disc is not eclipsed, as is the case in WZ Sge ($i = 76^{o}$). What about the bright spot in CG Dra? We note that our light curves show no evidence for a strong orbital hump, which is normally the signature of an active bright spot. Whether this is actually due to the lack of an active bight spot or to the bright spot being eclipsed is not certain. In either case, our observations are at least consistent with the flickering source in CG Dra being predominantly the inner disc.

To investigate the eclipse profile further, we combined 4 outburst eclipses (orbits 49, 53, 106, 116) by folding them on $P_{orb}$, having normalised the brightness. To reduce the impact of the flickering we plotted the block average of every 4 data points. However, the resulting average eclipse profile (Figure 7) shows an almost perfectly symmetrical eclipse, with no additional structure. We also took the same approach with three quiescence eclipse profiles (orbits 96, 97, 100) and found a similarly symmetrical profile (Figure 8) that was almost identical in shape and depth to the outburst eclipse profile (Δm = 0.16 mag). The symmetrical profile of the quiescence eclipse tends to imply that there is a single bright source - presumably the accretion disc - visible all the time. This would also explain why we don't see an orbital hump. We note that SW Sex stars (a class of “novalike” cataclysmic variable star, with $P_{orb}$ around 3 to 4 h and which exhibit eclipses [4, 16]) also have shallow symmetrical eclipses. However, these stars show high and low states rather than the dwarf nova outbursts that have been seen in CG Dra.

We suggest that further studies of the eclipses with a higher time resolution could be useful to look for additional structure in the eclipse profiles, especially relating to flickering. This would necessarily require larger telescopes than those used by the authors. Furthermore, our studies have only included observations during the “normal” range of outburst and quiescence states of CG Dra. It is known that brighter outbursts (up to ~mag 15.6) and fainter quiescence levels (to ~mag 17.2) occur from time to time [3]. It would therefore be interesting to compare the eclipses at these extreme levels, to look for changes in the eclipse profile and symmetry, which, for example, might reveal changes in the size of the accretion disc. Detailed studies of flickering during such eclipses may yield further information about the source of the flickering. For example does the flickering actually remain constant or is there a small reduction, caused by partial occultation of the source?

## Conclusion

Our results show for the first time that CG Dra is an eclipsing dwarf nova. By timing the minima of 13 eclipses, we derived an orbital period of 0.18864(4) d, or 4h 31m 38 +/- 3s. This value is consistent with one of two periods proposed from earlier spectroscopic studies. It places CG Dra well above the “period gap” in the orbital period distribution of dwarf novae. Given the very frequent outbursts and the lack of standstills, it is likely that CG Dra is a member of the UGSS family of dwarf novae.

The eclipses are symmetrical which means that the accretion disc is axisymmetric. They are of short duration (FWHM 18+/-2 min, or 0.066(7) of the orbital period) and shallow (average 0.16+/-0.02 mag) which is consistent with an orbital inclination just above the critical value. Moreover, the duration and depth of the eclipse is similar whether the star is in outburst or in quiescence. Flickering persists through the eclipse which means that the flickering source is not occulted by the secondary star.

## Acknowledgements

The authors gratefully appreciate the advice and encouragement given by Dr. Boris Gaensicke (University of Warwick) during this research. We acknowledge the use of SIMBAD, operated through the Centre de Données Astronomiques (Strasbourg). We are indebted to our referees for helpful comments which have improved the paper.

*Addresses:*
JS: “Pemberton”, School Lane, Bunbury, Tarporley, Cheshire, CW6 9NR
[bunburyobservatory@hotmail.com]

DB: 5 Silver Lane, West Challow, Wantage, Oxon, OX12 9TX, UK [drsboyd@dsl.pipex.com]
SB: 5 Melba Drive, Hudson, NH 03051, USA [sbrady10@verizon.net]
RP: 3 The Birches, Shobdon, Leominster, Herefordshire, HR6 9NG [rdp@astronomy.freeserve.co.uk]

| Observer | Telescope | CCD |
|---|---|---|
| JS | 0.28 m SCT | Starlight Xpress SXV-M7 |
| DB | 0.35 m SCT | Starlight Xpress SXV-H9 |
| SB | 0.4 m reflector | SBIG ST-8XME |
| RP | 0.30 m SCT | Starlight Xpress SXV-H9 |

**Table 1: Equipment used**

| Date in 2007 (UT) | Start time (JD-2454000) | Duration (h) | Number of eclipses seen | Observer |
|---|---|---|---|---|
| Sep 12 | 356.325 | 1.6 | 0 | RP |
| Sep 15 | 359.318 | 1.8 | 0[a] | RP |
| Sep 16 | 360.348 | 3.0 | 1 | RP |
| Sep 25 | 369.315 | 3.4 | 0 | JS |
| Sep 25 | 369.494 | 6.2 | 1 | SB |
| Sep 26 | 370.340 | 3.3 | 1 | JS |
| Sep 30 | 373.509 | 5.5 | 1 | SB |
| Oct 4 | 378.272 | 4.6 | 1 | JS |
| Oct 4 | 378.304 | 2.3 | 1 | DB |
| Oct 5 | 378.500 | 5.5 | 2 | SB |
| Oct 5 | 379.266 | 5.5 | 2 | JS |
| Oct 5 | 379.365 | 3.4 | 1 | RP |
| Oct 6 | 380.273 | 3.8 | 1 | JS |
| Oct 8 | 382.262 | 3.3 | 1 | JS |
| Dec 13 | 448.303 | 1.6 | 1 | DB |

**Table 2: Log of time-series observations**

[a] an eclipse egress was detected at the beginning of the run

| **Orbit no.** | **Eclipse time of min** (JD-2454000) | **Error** (d) | **Average mag. outside eclipse** | **Eclipse duration at FWHM** (min) | **Eclipse depth, Δm** (mag) |
|---|---|---|---|---|---|
| 0 | 360.4072 | 0.0014 | 16.5 | 16 | 0.18 |
| 49 | 369.6581 | 0.0011 | 16.1 | 18 | 0.16 |
| 53 | 370.4079 | 0.0016 | 16.0 | 19 | 0.18 |
| 70 | 373.6199 | 0.0017 | 16.5 | 20 | 0.13 |
| 95 | 378.3318 | 0.0014 | 16.7 | 19 | 0.21 |
| 95 | 378.3327 | 0.0016 | 16.7 | 17 | 0.19 |
| 96 | 378.5243 | 0.0013 | 16.7 | 18 | 0.14 |
| 97 | 378.7136 | 0.0017 | 16.7 | 18 | 0.16 |
| 100 | 379.2810 | 0.0015 | 16.6 | 15 | 0.15 |
| 100 | 379.2795 | 0.0011 | 16.6 | 17 | 0.15 |
| 101 | 379.4646 | 0.0012 | 16.6 | 18 | 0.16 |
| 106 | 380.4053 | 0.0015 | 16.0 | 19 | 0.15 |
| 116 | 382.2973 | 0.0013 | 16.2 | 16 | 0.16 |
| 466 | 448.3170 | 0.0014 | 16.8 | 17 | 0.17 |

**Table 3: Details of the eclipses**

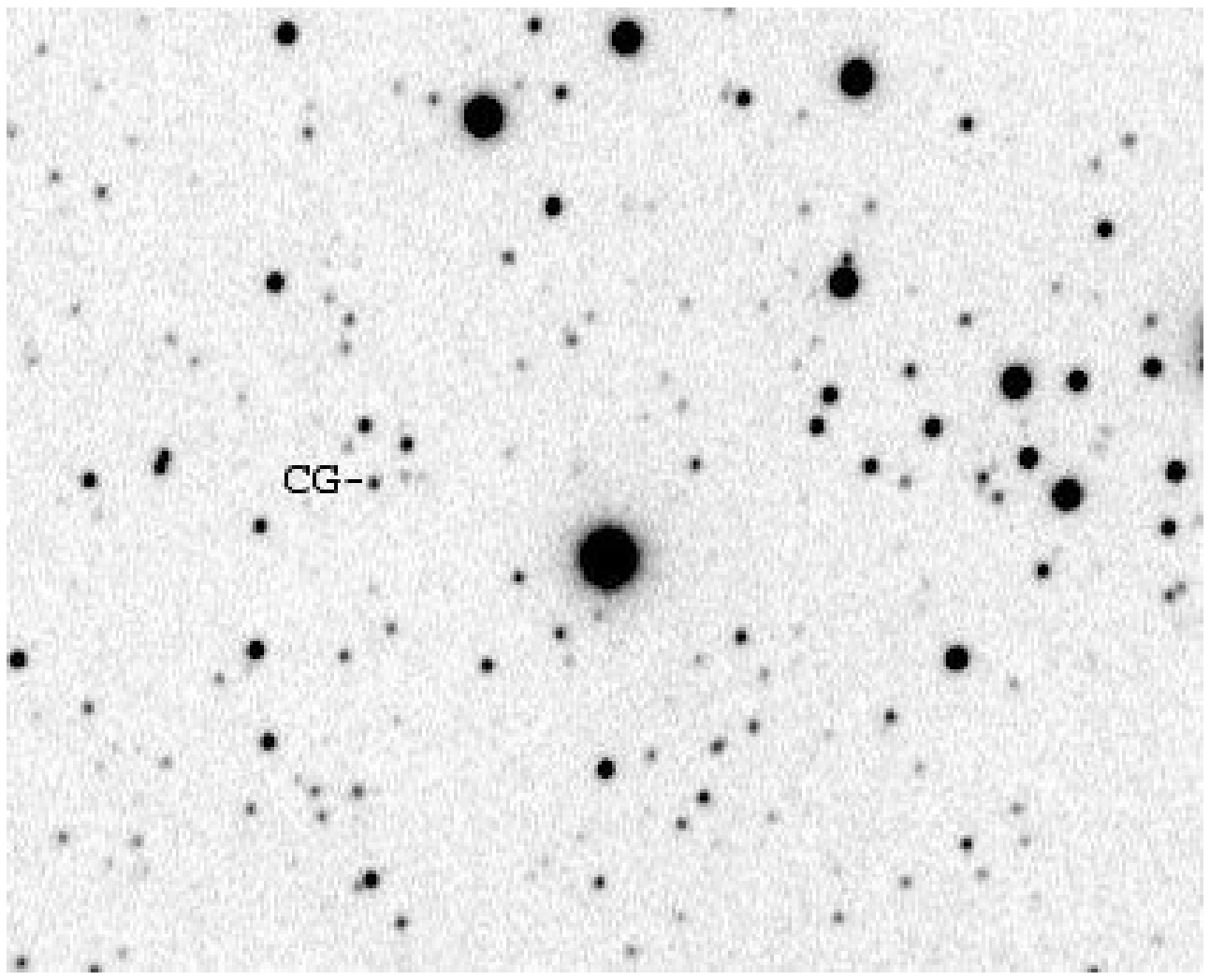

**Figure 1: CG Dra in outburst**
2005 April 9, 00.11UT. Takahashi FS102, 0.1 m refractor. 60 second image with unfiltered Starlight Xpress SXV-M7 CCD. Field 11.5’ x 9’ with South at the top and East to the right

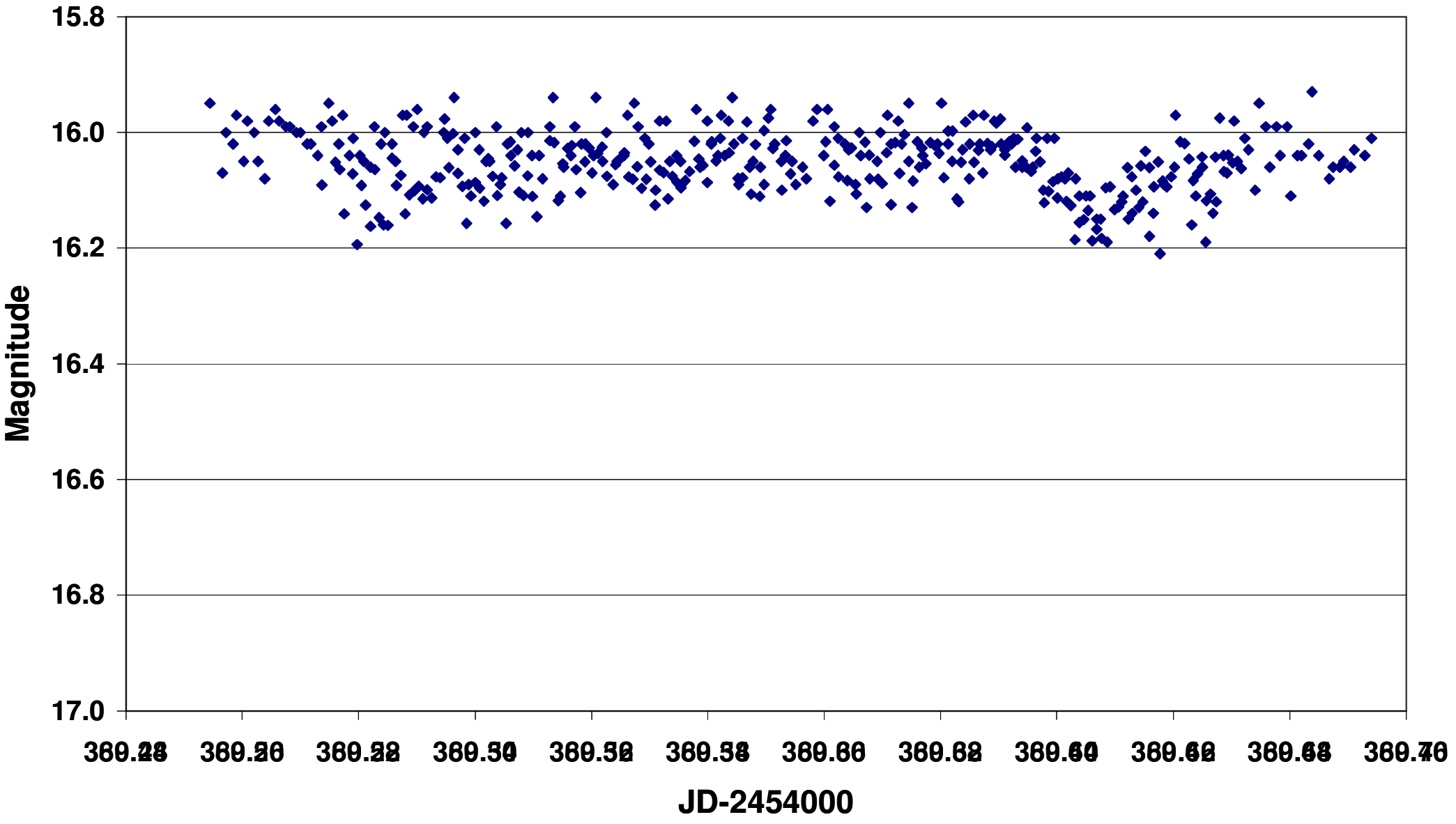

**Figure 2: 2007 Sep 25 (JD 2454369) – in outburst**

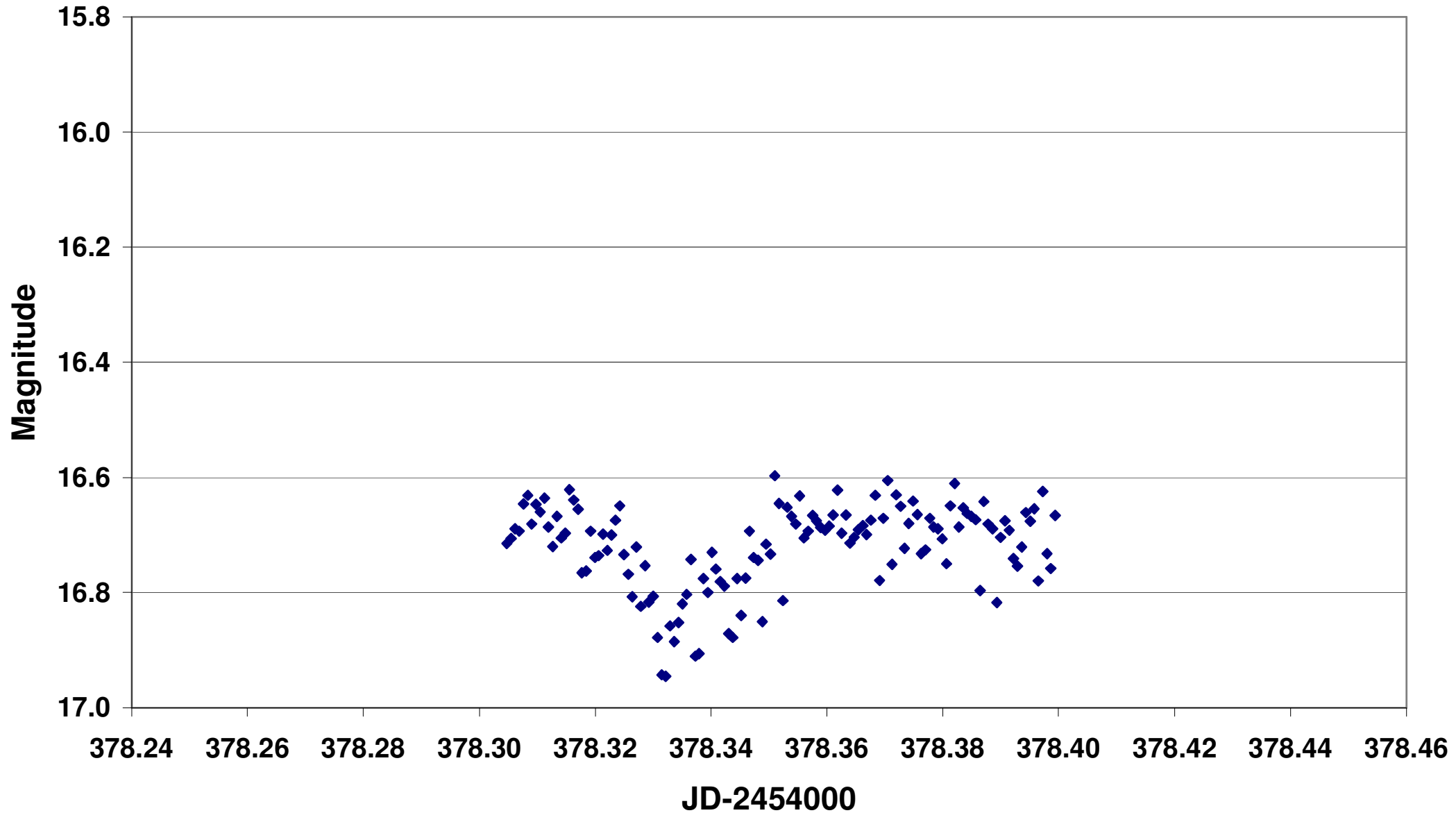

**Figure 3: 2007 Oct 4 (JD 2454378) – near quiescence**

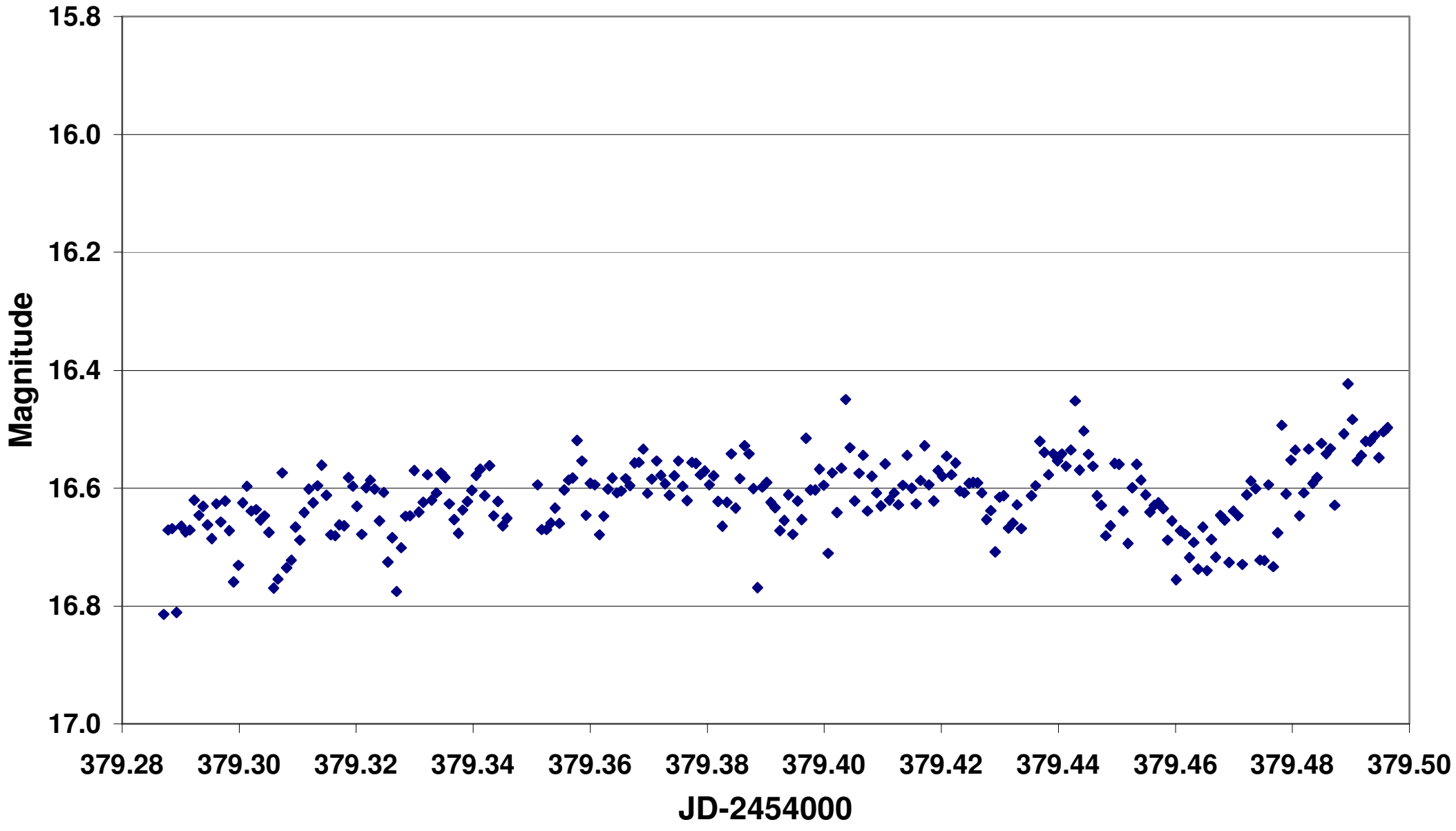

**Figure 4: 2007 Oct 5 (JD 2454379) – near quiescence**

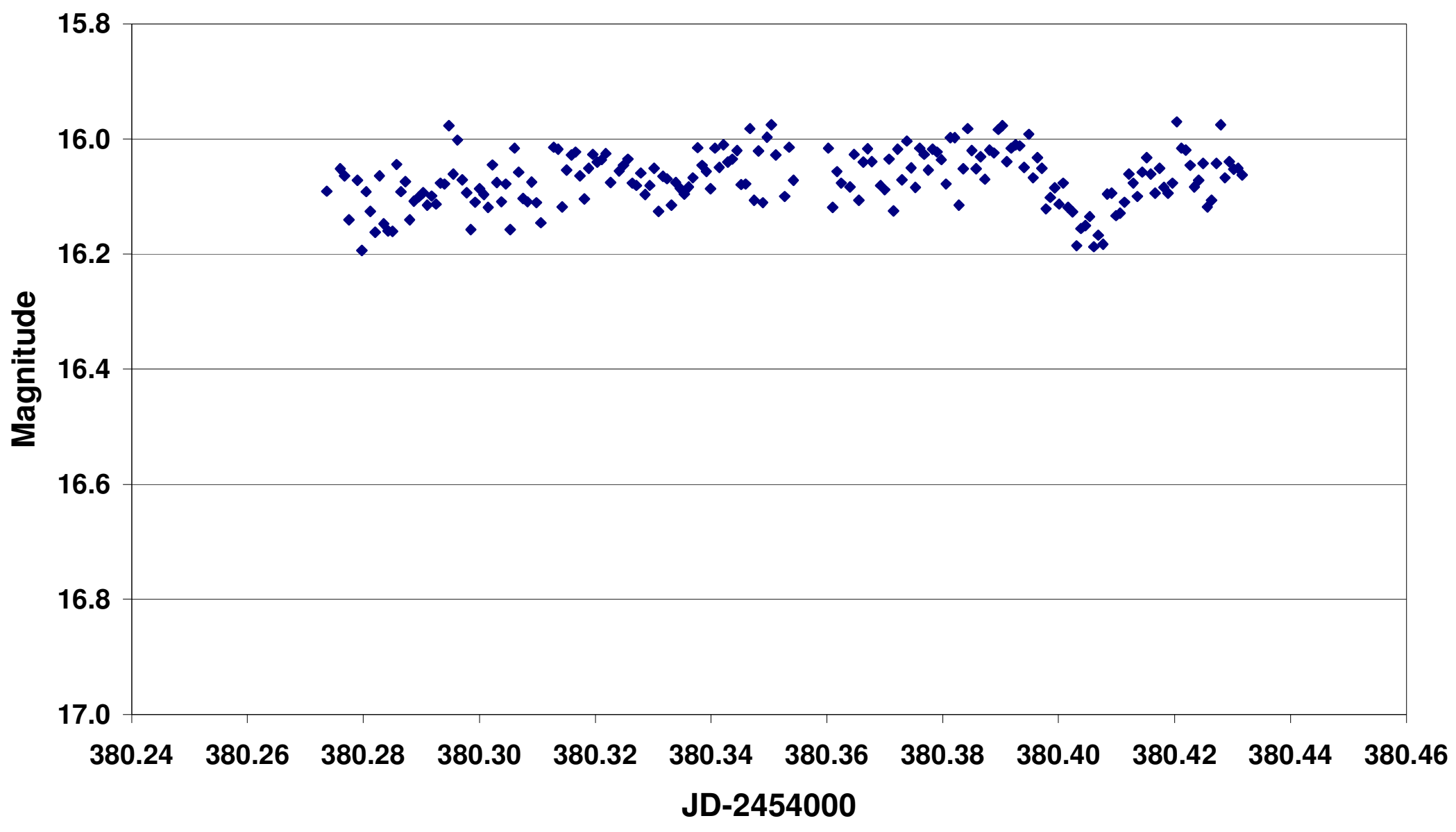

**Figure 5: 2007 Oct 6 (JD 2454380) – in outburst**

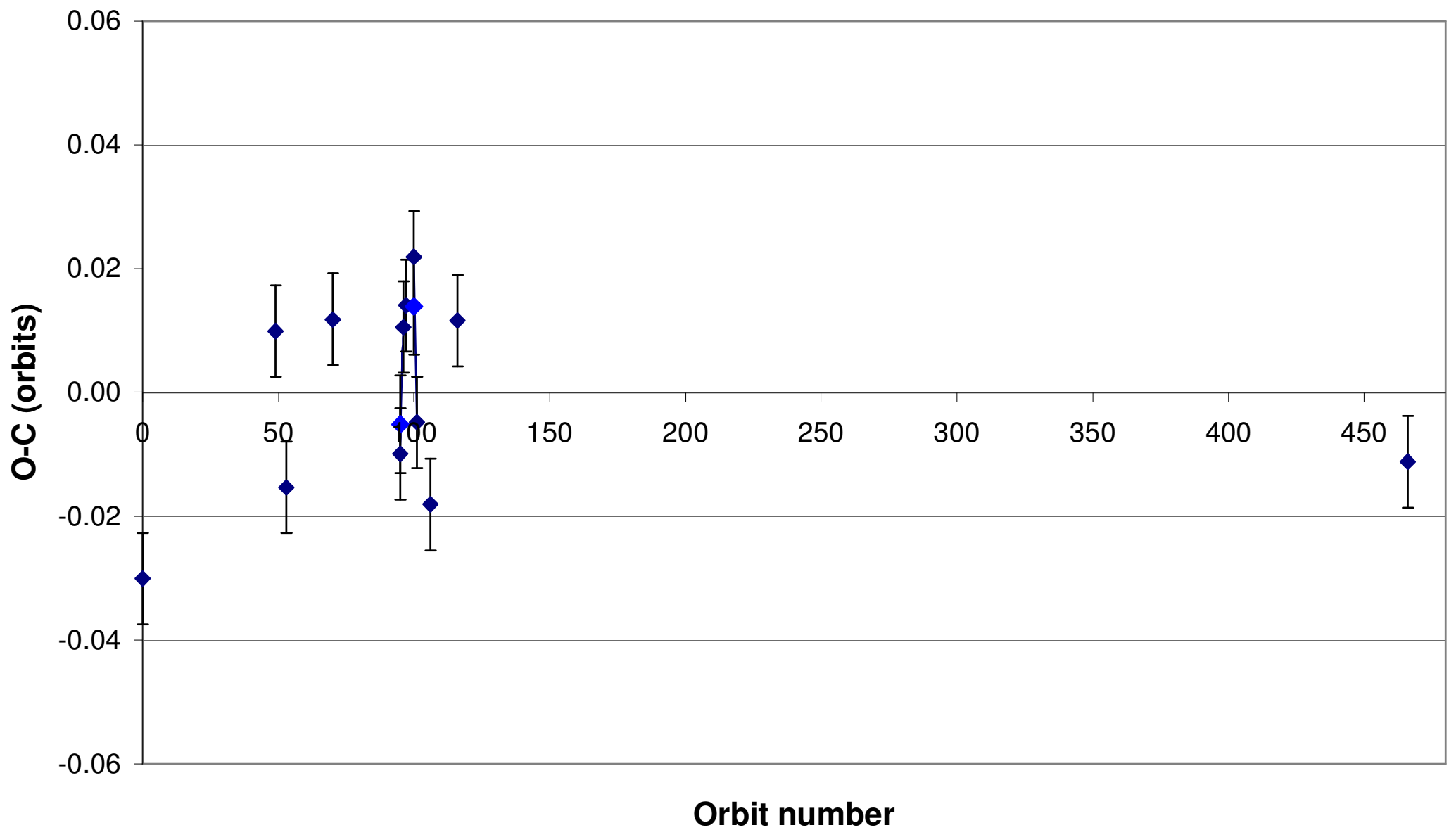

**Figure 6: O-C residuals for the eclipses**

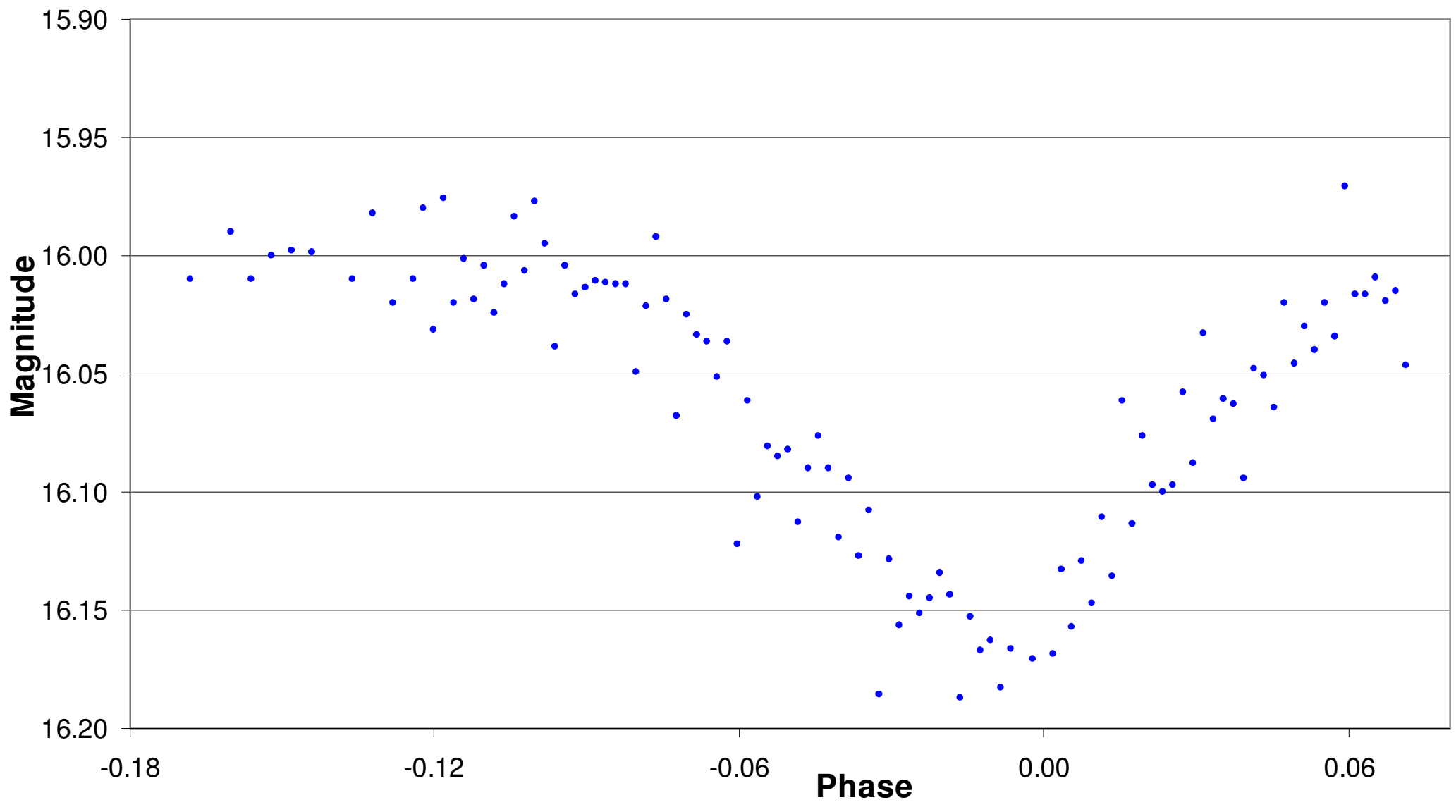

**Figure 7: Average outburst eclipse profile of 4 eclipses obtained by folding eclipses onto $P_{orb}$**

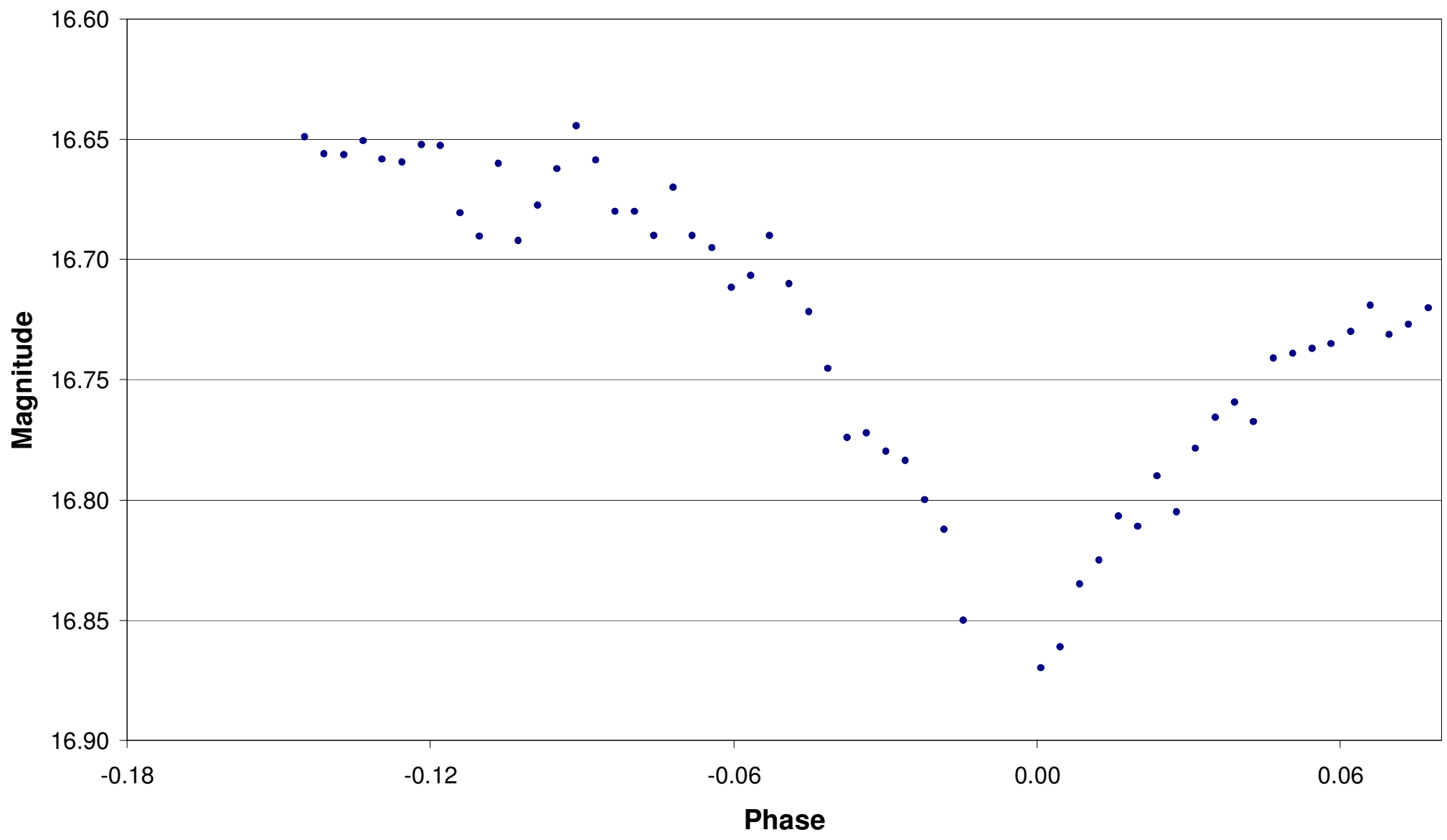

**Figure 8: Average quiescence eclipse profile of 4 eclipses obtained by folding eclipses onto $P_{orb}$**